\documentstyle[preprint,aps]{revtex}
\begin{document}
\draft
\title
 {High temperature  study of the Kosterlitz-Thouless phase\\ 
transition in the  XY model  on the triangular lattice\\} 
\author{P. Butera and M. Comi}
\address
{Istituto Nazionale di Fisica Nucleare\\
Dipartimento di Fisica, Universit\`a di Milano\\ 
Via Celoria 16, 20133 Milano, Italy}
\maketitle
\begin{abstract}
Abstract: High temperature series 
expansions of the spin-spin correlation
function for the XY (or plane rotator) model 
on the triangular lattice are  extended 
by two terms up to  order
$\beta^{14}$. Tables of the expansion coefficients 
are reported for the correlation function
spherical moments of order $l=0$ and $2$.
Our analysis of the 
series supports the 
Kosterlitz-Thouless predictions on 
the structure of the critical
singularities and leads to  fairly 
accurate estimates of the 
critical parameters.
\end{abstract}
\section{ Introduction }

In the last two decades since the seminal papers
 by Berezinski and by Kosterlitz and Thouless,
the critical  behavior of the 
two-dimensional XY (or plane rotator) model 
 has  been much studied numerically,  
 mainly on the square lattice 
\cite{htstudies,httria,hamer,oldmc,ferervel,seiler}
\cite{bcm,bc,gupta,wolff,edwards,janke,biferale}, 
 but, sometimes, 
  also on the triangular lattice \cite{httria,ferervel}.
The steady  increase of the computers power,
   important recent  progress in 
  MonteCarlo (MC) algorithms \cite{sokal}
and the calculation  of  $O(\beta^{20})$ high 
temperature expansions (HTEs)  on  the square lattice \cite{bc},
 have  produced 
increasingly accurate verifications of 
the Kosterlitz and Thouless (KT) 
theory \cite{kt,bere}.
 However, since on reasonable grounds 
 it has been disputed \cite{seiler,edwards}
that these studies are really conclusive,
further quantitative evidence is still valuable.
 
Here we present  an extension (by two terms 
up to order $\beta^{14}$)
and a new analysis, by the methods of Ref.\cite{bc},
 of HTEs for the XY model on the triangular lattice.
Our study gives further  support to the KT picture and 
leads to fairly precise 
estimates of the KT critical parameters. On the contrary 
our results appear not to be consistent with
 a conventional power-law critical behavior unless 
the exponents are very large,
 namely $\gamma \gtrsim 3.2 $ and $\gamma+2\nu \gtrsim 5.4 $. 
If  this is the case, our $O(\beta^{14})$
 series do not seem to be sufficiently long
to yeld accurate estimates of the critical exponents.

In the second Section, we  recall briefly the main 
 predictions of the KT theory. 
The third Section is devoted
to an analysis of the series by ratio-extrapolation,
and by rational and differential approximants techniques.
The last Section contains  our conclusions.

\section{ The model and the HT series}
\label{sec:ht}

The Hamiltonian  of the two-dimensional 
 XY model is
\begin{equation}
H \{ s \} = - \frac{1} {2} \sum_{\langle x,x' \rangle } s(x) \cdot s(x').
\label{hamilt} \end{equation}
Here $s(x)$ is a two-component classical spin 
of unit length associated
to the site  with position vector
$ x $ of a 2-dimensional triangular
lattice
 and the sum extends over all nearest neighbor 
lattice pairs $\langle x,x' \rangle $.  

Our series  have been computed by  a FORTRAN code 
which solves recursively the Schwinger-Dyson 
equations for the correlation  functions \cite{bcm,guerra}.

Here we shall analyze  
 the HTE of the spherical moments of
the correlation function  $m^{(l)}(\beta)$
 for $l=0$ ( reduced susceptibility ) and  $l=2$.

Table I shows the HTE 
coefficients through $\beta^{14}$
of the nearest-neighbor spin-spin correlation function.
In Tables II and III we have reported the 
expansion coefficients for the moments 
$ m^{(l)}(\beta) $ 
with $ l= 0$ and $2$.

The main predictions of 
the non-rigorous renormalization 
group analysis of the plane rotator model \cite{kt,bere}
 can be summarized as follows.

The correlation length $\xi(\beta)$ 
is expected to diverge as 
$ \beta \uparrow \beta_{c}$ 
with the unusual singularity
\begin{equation}
\xi(\beta) \propto \xi_{as}(\beta)=  exp(\frac {b} 
{\tau^{\sigma}})[ 1 + O(\tau)] 
\label{corleng} \end{equation}
where $\tau=\beta_{c}-\beta$.

The value of the exponent $ \sigma$ 
predicted in Ref.\cite{kt}  is 
 $ \sigma=1/2 $ and $b$ is a non-universal 
positive constant.

At the critical temperature, the 
asymptotic behavior of the two-spin
correlation function as 
$ r=|x| \rightarrow \infty$ is expected to be
\begin{equation}
<s(0)\cdot s(x)> \propto 
\frac{({\rm ln}(r))^{2\theta}} {r^{\eta}}
[1+O({\rm ln}({\rm ln}(r))/{\rm ln}(r))]
\label{asycor} \end{equation}
The values predicted \cite{kt,amit}  for $ \eta$   and 
$ \theta$ are, respectively, $ \eta=1/4 $ , 
$ \theta=1/16 $.

From  Eqs. (\ref{corleng}) 
and (\ref{asycor}), together with the usual scaling ansatz, 
it follows that,  
for $ l > \eta-2 $,  the 
 correlation moment $m^{(l)}(\beta)$ should
diverge as 
$ \beta \uparrow \beta_{c}$ with the singularity
\begin{eqnarray}
m^{(l)}(\beta) \propto
\tau^{-\theta}\xi_{as}(\beta)^{2-\eta+l}
[ 1 + O(\tau^{\frac {1} {2}}
{\rm ln}(\tau))] 
\label{asymom} \end{eqnarray}

At $\beta_{c}$  a line of critical
points should begin which extends to $ \beta=\infty$, 
so that for $ \beta >\beta_{c}$
both $ \xi$ and the correlation moments remain infinite.

Finally it is worth mentioning the rigorous bound \cite{aize}
\begin{eqnarray}
\beta_c \geq 2 \beta_c^I = \frac{1} {2} ln3 = 0.5493... 
\label{bound}
\end{eqnarray}
where $ \beta_c^I $ is the inverse critical temperature of the Ising
 model on the triangular lattice.
 
\section{ An analysis of the HT series }
\label{sec:analys}
 We  estimate the critical parameters  by 
simple modifications of  current 
methods of series analysis \cite{bcm,bc,tonyg,seran,nickel,diffapp}.

In Refs. \cite{bcm,bc} we  have shown
that the ratios 
$r_n(m^{(l)})= a^{(l)}_n/a^{(l)}_{n+1}$ 
of the successive HTE coefficients of
 the  correlation moment $ m^{(l)}(\beta)$
for large $n$ behave as:
\begin{equation}
 r_{n}(m^{(l)})=\beta_{c}+\frac {C_{l}} 
{(n+1)^{\epsilon}}  +O(1/n^{\lambda})
\label{ratiokt} \end{equation}
with $\epsilon=\frac {1} {1+\sigma} $, 
$C_{l}=-((2-\eta+l) \sigma b  \beta_{c})^{\epsilon}$
 and $\lambda = min(1,2\epsilon)$.
 According to
the KT prediction we should have $ \epsilon=2/3 $. This
 is a clear
 signature of the KT singularity and it should
 be detected in a ratio analysis of the HTEs. 

 On the other hand if, instead of (\ref{corleng}) 
and (\ref{asymom}), we had  
conventional power-law critical behavior so that, 
 as $ \beta \uparrow \beta_{c}$, 
\begin{equation}
m^{(l)}(\beta) \sim  
\tau^{-\gamma -l\nu} 
[ A_{l} + B_{l} \tau^{\Delta}+..] 
\label{asymompow} \end{equation}
where, $\Delta > 0$ and, as usual, 
$\gamma$ and $\nu$ denote 
the susceptibility and the 
correlation length exponents respectively,
we would obtain a formula analogous 
to Eq.\  (\ref{ratiokt})
with $\epsilon=1$ and ${\lambda = 1+\Delta}$,
namely
\begin{equation}
 r_{n}(m^{(l)})=\beta_{c}+\frac {\beta_{c}(1-\gamma-l\nu)} {n}  
+O(1/n^{ 1+\Delta})
\label{ratiopow} \end{equation}

In Fig.1 we have plotted versus  $1/n$ the 
sequences of  ratios
$ r_{n}(m^{(l)})$ for $l=0$ and $2$.
 These ratio plots exhibit a 
residual curvature and an increasing slope for large $n$.
If Eq.\ (\ref{ratiopow}) were a correct 
representation of the asymptotic
behavior of $ r_{n}(m^{(l)})$, 
the $O(1/n)$ terms in (\ref{ratiopow}) 
 should be  suppressed by forming the 
linearly extrapolated sequences
\begin{eqnarray}
r^{(1)}_{n}(m^{(l)})= n r_{n}(m^{(l)})-
(n-1) r_{n-1}(m^{(l)})
=\beta_{c} + O(1/n^{ 1+\Delta})
\label{extrapolin} \end{eqnarray}
which, for large $n$,  should approach 
with vanishing slope their common
limit $\beta_{c}$.
This does not  happen (at least up to $n=14$),  as it is  shown
in Fig.1 where we have also plotted the 
 sequences $ r^{(1)}_{n}(m^{(l)})$ 
 versus $1/n$. The estimates of $\beta_{c}$ obtained
 from $r^{(1)}_{n}(m^{(l)})$ still  increase rapidly with order.

Turning to critical exponents, we have  computed a 
sequence of (unbiased) estimates of 
$\gamma+ l\nu$ by the formula 
\begin{equation}
(\gamma+ l\nu)_{n}= 
\frac {(n-1)^2 r_{n}(m^{(l)})-n(n-2) r_{n-1}(m^{(l)})}
{n r_{n-1}(m^{(l)})-(n-1) r_{n}(m^{(l)})}
\label{gamplusnu} \end{equation}

The sequences of 
estimates so obtained for $\gamma$ and $\gamma +2\nu$
 are plotted versus $1/n$ in Fig.2.
We conclude
 that,
 under the assumption of 
power-law scaling, 
the simplest extrapolations  suggest 
$\gamma \gtrsim 3.2 $ and $\gamma+2\nu \gtrsim 5.4 $. 
These estimates for the exponents are larger than those 
 obtained from a fit \cite{seiler}
of ( square lattice) MC data  to power law critical behavior.

Let us assume now that Eq.\ (\ref{ratiokt})
is valid instead of Eq.\ (\ref{ratiopow}), then by 
reporting the $  r_{n}(m^{(l)})$ sequences
versus $1/n^{2/3}$, we should obtain nicely straight plots,
as indeed is observed in Fig.3. 
We can 
suppress the $O(1/n^{2/3})$ terms
in the sequences $  r_{n}(m^{(l)})$
 by forming the (nonlinearly) extrapolated sequences
\begin{eqnarray}
 s_{n}(m^{(l)})=
\frac{n^{2/3} r_{n}
(m^{(l)})-(n-1)^{2/3} 
 r_{n-2}(m^{(l)})} {n^{2/3}-(n-1)^{2/3}}
=\beta_{c}+ O(1/n)
\label{extrapokt} \end{eqnarray}
Unfortunately the sequences  obtained are not regular enough 
to give a much  more precise estimate of $\beta_{c}$
by a further extrapolation in $1/n$. 
The results are reported in Fig.3. 
 We can infer that $\beta_{c}=0.683 \pm0.004$. 

 A direct unbiased estimate of   
$\epsilon $ in terms of ratios is obtained 
 from the sequence
\begin{equation}
\epsilon_{n}=n {\rm ln}( \frac {t_n -1} {t_{n+1}-1} ) 
\label{enp} 
\end{equation}
where 
$t_n  = \frac { r_{n}( \chi^2 )} { r_{n}(\chi)}$.
 If the ratios 
$ r_{n}(\chi)$ and $ r_{n}(\chi^2)$  
have the asymptotic behavior 
 (\ref{ratiokt}), the
sequence $\epsilon_{n}$
 will provide estimates
of  $\epsilon$.
A quantity $u_n$ 
analogous to $t_n$ may be defined in 
terms of the moment $m^{(2)}(\beta)$  
and its square, 
and, via  Eq. \ref{enp}, the corresponding 
sequence $\epsilon_{n}'$  may be formed.
 The sequences  $\epsilon_{n}$ and $\epsilon_{n}'$ 
have been plotted versus $ 1/n $  in Fig.4.
We have also reported the 
 sequence $\bar \epsilon_{n}$ as computed from the
 susceptibility $\chi_I$  of the triangular lattice 
Ising model \cite{sykes},
in order to emphasize the qualitatively 
different behavior of the two cases.
These tests  support  the KT 
theory and are inconsistent with a power-law singularity: 
 if that were  the case, as it appears from the  Ising model plot,
 the limit of the sequences should be 1. 
An extrapolation of  
 $\epsilon_{n}$ and $\epsilon_{n}'$  to $n=\infty$
 by the Barber-Hamer method \cite{tonyg} leads to the estimate
$\sigma= 0.51 \pm 0.04$, while for the sequence $\bar \epsilon_{n}$ we get
 $\sigma= 0.02 \pm 0.04$.
Notice that this test is able to distinguish 
sharply  power-like from 
KT singularities, unless asymptotic 
behaviors have not yet set in.
 This is true even if the critical 
exponents are  large: 
 for instance a reasonable 
 model series, like $\chi_I^2$
 ( having $\gamma=3.5$), produces an $\epsilon$-sequence with the same
qualitative behavior as $\bar \epsilon_{n}$.

The rest of our analysis uses differential approximants (DAs) 
or simply Pad\'e approximants (PAs).

The expected singularity structure of $ ln(\chi)$, namely
$ ln(\chi) = A(\beta)(\beta_c-\beta)^{-\sigma} +B(\beta) $ 
with $A(\beta)$  regular and $B(\beta)$   
 at most weakly singular at $\beta_c$, should be 
reproduced with reasonable accuracy by inhomogeneous first order DAs.
We have selected DAs $[n/l;m]$ with 
$ 1 \le  n \le 6 $, $ 2 \le  m \le 5 $ and $ 2 \le l \le 6$.
From the non-defective approximants of the 
reduced sample which use at least 12 series coefficients, 
 we get the unbiased estimates $\beta_c = 0.680 \pm 0.002$
and $\sigma = 0.49 \pm 0.03$.
If we set $\sigma = \frac {1} {2} $, we get the biased estimate
$\beta_c = 0.681 \pm 0.002$. 
Conservatively we have estimated uncertainties as
 three times the standard deviation of the sample.
The results remain essentially unchanged whether we
use the Fisher-Au Yang-Hunter-Baker or the Guttmann-Joyce
definitions \cite{tonyg} of the DAs.

 We have also computed the PAs to 
the logarithmic derivative of ${\rm ln}(\chi)/\beta$. 
This quantity should   discriminate between the 
structures (\ref{asymom}) and (\ref{asymompow}) 
of the critical singularity, 
since the residues at the critical poles 
have either to approach  $ \sigma $, if (\ref{asymom})
holds, or to vanish, if (\ref{asymompow})  holds.
 From the PA table for the 
location of the critical pole of
the approximants to
$ D{\rm ln}[ {\rm ln}(\chi)/\beta] $ and 
 the PA table for the residues   
 we get  the estimates
 $\beta_{c} = 0.684 \pm 0.003$ and 
 $\sigma = 0.54 \pm 0.04$.
If we set $\sigma = \frac {1} {2} $, we get the biased estimate
$\beta_c = 0.680 \pm 0.003$. 
 The PA and the DA estimates are  therefore consistent, 
the small difference in the central values being probably
due to  background effects.

Due to the slower convergence of the $m^{(2)}$ series a 
 similar DA analysis  of the correlation length 
(using  ${\rm ln}(\frac {\xi^{2}} {\beta })$ )  
 is unsuccessful unless we assume
$\sigma = \frac {1} {2} $. In this case we get the estimate
$ \beta_c = 0.684 \pm 0.004 $
consistent with the previous ones from $\chi$,
 but somewhat less accurate.
 
Assuming $\sigma = \frac {1} {2} $,  the non-universal parameter $  b $ 
may be obtained by computing PAs of the quantity
\begin{equation}
 C(\beta)= \frac{1} {2} (\tau)^{\sigma} {\rm ln}(1+m^{(2)}/\chi)=
b +O(\tau^{\sigma})
\label{b} \end{equation}
 at $\beta = \beta_c$.
Taking $\beta_c$ in the range $[0.680 , 0.682]$,
we estimate $b = 1.27 \pm 0.05$.

Finally, the critical index  $\eta$
governing the large distance behavior 
of spin-spin correlation 
function may be estimated (without bias on $\sigma$) 
 by PAs of the ratio 
\begin{equation}
 H(\beta)= \frac  {{\rm ln}(1+m^{(2)}/\chi^2)} {{\rm ln}(\chi)}=
\frac {\eta} {2-\eta} +O(\tau^{\sigma})
\label{eta1} \end{equation}
 at $\beta = \beta_c$.
Allowing as above for the uncertainty on 
$\beta_c$, we estimate $\eta= 0.27 \pm 0.01$.

If we assume a power-law singularity 
 (\ref{asymompow}), from a study 
 of the location of the 
singularities of the PAs to the logarithmic 
derivative of the susceptibility 
$D{\rm ln}(\chi)$, we should be able to estimate $\beta_{c}$,
and from their residues, the critical exponent $\gamma$.
As we have already observed in the 
case of the HTEs for the square lattice
\cite{bc}, both the PA tables 
for the poles and for 
the residues contain many "defective entries" 
or blanks and  show  a  very poor
convergence. These features of the 
PAs   suggest that the critical
singularity is not a power or, at least,
that our series are still too short.
If we insist  in producing anyway  some estimate 
of the critical parameters,
then, by averaging over all relevant entries 
 of the PA tables 
for the poles and residues of 
the approximants to $D{\rm ln}(\chi)$
we get  $\beta_{c} = 0.655 \pm 0.008$
and $\gamma = 3.7 \pm 0.6$. 
This estimate for $\gamma$ is  consistent with those from
ratio tests,  
 but larger than those obtained in Ref.\cite{seiler} from 
 power-law fits to MC data.

\section{ Conclusions }
\label{sec:conc}
Let us finally compare our results to those 
obtained in previous studies of the model
 on the triangular lattice.
To our knowledge, no MC simulation  
is available for the ferromagnetic XY model
 on the triangular lattice although, by 
virtue of the higher coordination number of this lattice, 
the approach to scaling behavior is expected to be 
 faster than in the 
square lattice case, for the same lattice size.
On the same grounds one can also argue \cite{tgut} 
that, for a given number of HT coefficients, 
the triangular lattice series 
are "effectively longer" than the square 
lattice ones. In fact, we expect  from our 
analysis that it would perhaps 
take only an O($\beta^{16}$)
 triangular lattice series  to reach 
the same precision as with our
O($\beta^{20}$) square lattice series \cite{bc}.

 An extensive review of the square lattice 
 numerical studies can be found in Refs. \cite{bc,edwards}  
 and needs no  duplication here.
 As to HT studies \cite{htstudies}  
 on the triangular lattice, we recall that the early ones, 
  based on ten term series, 
were  essentially inconclusive 
 and  did not  provide  reasonably stable estimates
of the critical parameters. 
 When  the HT series were  extended 
 to twelve terms \cite{ferervel}, 
 an analysis by the four-fit method  gave  convincing
indications of the validity of the KT predictions
 and yelded the  estimates 
 $\beta_{c}=0.687 \pm 0.009$,  
$\sigma=0.5 \pm 0.1 $ and $\eta= 0.27 \pm 0.03$. 

We believe that our extended series and our new tests
 further substantiate the  KT picture
and  provide more precise estimates of the critical parameters,
 otherwise  consistent with those of Ref. \cite{ferervel}.
We have  pointed out
 that the study of ratio plots, and of
the PAs to the logarithmic derivative of $\chi$ and $\xi$, 
 are strongly suggestive that the critical singularity is not a power.
If, however, we want to pursue  a power-law interpretation, 
we should not overlook the unusual facts that, 
for a closely packed lattice, $O(\beta^{14})$ series
do not seem long enough to determine with reasonable accuracy the
  critical parameters $\gamma$ and $\nu$ and that the 
rough estimates obtained for these exponents are  
 significantly larger than those   observed in conventional 
critical phenomena and show a  trend to increase
with the number of  coefficients used.
These estimates are also larger than those
obtained in  the power-law fit of Ref. \cite{seiler} 
to square lattice MC data.
 However we have  reexamined the most recent 
and extensive MC simulation data \cite{gupta,edwards}
on the square lattice and we have observed  that they 
are consistent with a power-law finite size scaling Ansatz
if we choose the larger exponents
suggested by our series ( it seems that this range of parameters 
has not been explored in the fits of Refs. \cite{gupta,edwards}).
Our interpretation of all these facts 
has been already pointed out at the conclusion  of our analysis of the square 
lattice series \cite{bc}: as the critical point is approached, 
 larger and larger effective exponents are needed to fit  an infinite 
order singularity by an ordinary power singularity.

 We may  conclude that, if we
 extrapolate the HT series  consistently with the
KT behavior, we find stable values of the critical parameters 
in good agreement with the KT fits to the MC
data and with our previous studies of HTEs on square lattice. 
 Our  unbiased estimates
 of the critical parameters are 
 $\beta_{c}=0.680 \pm 0.002$,
$\sigma=0.49 \pm 0.03 $ and $\eta= 0.27 \pm 0.01$.

If we set  $\sigma=\frac{1} {2} $ we obtain the  estimate 
 $\beta_{c}=0.681 \pm 0.002$ and by also fixing 
$\beta_{c}$ at its central value 
 we find  $b=1.27 \pm 0.05$.

\section*{Acknowledgments}
\label{ack}
Our work has been partially supported by MURST.

\begin{figure}
\caption{The successive ratios of the HTE coefficients 
of various moments are plotted versus $1/n$. 
 The  ratios $  r_{n}(\chi) $ are represented by
solid squares; $ r_{n}(m^{(2)}) $ by solid triangles.
 We  have also plotted the  linearly extrapolated ratio sequences
$  r_{n}^{(1)}(\chi) $ (open squares), and  
$ r_{n}^{(1)}(m^{(2)}) $  (open triangles).}
\label{ratioplot}
\end{figure}

\begin{figure}
\caption{ Unbiased estimates of the critical exponent $\gamma$ of
the susceptibility under the assumption of a power-law critical 
singularity obtained from the  ratios $  r_{n}(\chi) $
(open squares). Analogous estimates of the exponent $\gamma + 2 \nu$
 as obtained from $ r_{n}(m^{(2)}) $ (open triangles).}
\label{gammapower}
\end{figure}

\begin{figure}
\caption{Ratio plots for the HTE coefficients versus $1/n^{2/3}$. 
 The  ratios $  r_n(\chi) $ are represented by
solid squares; $ r_n(m^{(2)}) $ by solid triangles.
The  ratio sequences  have been 
 extrapolated in $1/n^{2/3}$ obtaining the sequences
$  s_n(\chi) $ (open squares), $ s_n(m^{(2)}) $  
(open triangles). A further extrapolation in $1/n$
of the sequences $s_n$ gives $  s_n^{(1)}(\chi) $ 
(open circles), $ s_n^{(1)}(m^{(2)}) $  (crosses).}
\label{ratioplott}
\end{figure}

\begin{figure}
\caption{ The sequences $\epsilon_n$ (open squares), 
 $\epsilon_n'$ (open triangles), as computed from the quantities $t_n$ 
introduced in Eq. \protect\ref{enp} , and from the analogous ones $ u_n $,  
are plotted  versus $1/n$. The dashed line indicates the KT prediction 
for the  value of $ \epsilon $.  We have also reported
the analogous sequence $\bar \epsilon_n $ (crosses) as computed from
the susceptibility of the triangular lattice Ising model.}
\label{epsequence}
 \end{figure}

\narrowtext
\begin{table}
\caption{HTE coefficients of the 
nearest neighbor  correlation  function.}
\begin{tabular}{cc}
order&coefficient\\
\tableline
          1&   0.500000000000000000000000000000\\      
          2&   0.500000000000000000000000000000\\      
          3&   0.437500000000000000000000000000\\      
          4&   0.312500000000000000000000000000\\      
          5&   0.182291666666666666666666666667\\      
          6&   0.072916666666666666666666666667\\
          7&  -0.039550781250000000000000000000\\
          8&  -0.204481336805555555555555555556\\      
          9&  -0.448027886284722222222222222222\\      
         10&  -0.763769531250000000000000000000\\      
         11&  -1.139343883373119212962962962962\\      
         12&  -1.581509003815827546296296296296\\      
         13&  -2.114185611785404265873015873016\\      
         14&  -2.760278320043385554453262786596\\
\end{tabular}
\label{Cnn}
\end{table}
\narrowtext
\begin{table}
\caption{HTE coefficients of the 
susceptibility $m^{(0)}$.}
\begin{tabular}{cc}
order&coefficient\\
\tableline
 0    &1.000000000000000000000000\\
 1    &3.000000000000000000000000\\
 2    &7.500000000000000000000000\\
 3    &16.87500000000000000000000\\
 4    &35.62500000000000000000000\\
 5    &72.06250000000000000000000\\
 6    &141.2734375000000000000000\\
 7    &270.1728515625000000000000\\
 8    &506.3834635416666666666666\\
 9    &933.5703776041666666666666\\
10    &1697.512101236979166666666\\
11    &3050.264278496636284722222\\
12    &5424.862119886610243055555\\
13    &9561.162654477074032738095\\
14    &16716.55156094636866655299\\
\end{tabular}
\label{mom0}
\end{table}
\narrowtext
\begin{table}
\caption{HTE coefficients of the 
second correlation moment $m^{(2)}$.}
\begin{tabular}{cc}
order&coefficient\\
 0   &0.000000000000000000000000000\\
 1   &3.000000000000000000000000000\\
 2   &18.00000000000000000000000000\\
 3   &72.37500000000000000000000000\\
 4   &239.6250000000000000000000000\\
 5   &703.7500000000000000000000000\\
 6   &1902.406250000000000000000000\\
 7   &4835.969726562500000000000000\\
 8   &11719.07975260416666666666666\\
 9   &27326.64085286458333333333333\\
10   &61726.45278320312500000000000\\
11   &135743.4609174940321180555555\\
12   &291741.0660864935980902777777\\
13   &614640.8839340452163938492063\\
14   &1272465.830598210733403604497\\
\end{tabular}
\label{mom2}
\end{table}

\begin{references}
\bibitem{htstudies} H. E. Stanley, T. A. Kaplan
Phys.\ Rev.\ Lett. {\bf 17}, 913 (1966); H. E. Stanley, 
Phys.\ Rev.\ Lett. {\bf 20}, 589 (1968);
 M. A. Moore, Phys.\ Rev.\ Lett. 
{\bf 23}, 861 (1969); 
 D. N. Lambeth and H. E. Stanley, 
Phys. \ Rev.\ B{\bf 12} 5302 (1975).
\bibitem{httria} W. J. Camp, 
J. P. Van Dyke, J. \ Phys. \ C{\bf 8},  336 (1975); 
 A. J. Guttmann, J. \ Phys. \ A{\bf 11}, 545 (1978). 
\bibitem{hamer} C. J. Hamer and J. B. Kogut, 
Phys. \ Rev.\ B{\bf 20}, 3859 (1979);
 C. J. Hamer and M. N. Barber, 
J. \ Phys.\ A{\bf 14},  259 (1981);
 H. W. Hamber and J. Richardson, 
Phys. \ Rev.\ B{\bf 23}, 4698 (1981);
 M. Kolb, R Jullien and P. Pfeuty, 
J.\ Phys.\ A{\bf 15}, 3799 (1982);
 P.G.Hornby and M.N. Barber, 
J.\ Phys.\ A{\bf 18}, 827 (1985);
 C. R. Allton and C. J. Hamer, 
J.\ Phys.\ A{\bf 21}, 2417 (1988).
\bibitem{oldmc}M. Suzuki, S. Miyashita, 
A. Kuroda, C. Kawabata, 
 Phys. \ Lett.\ B{\bf 60}, 478 (1977);
 C. Kawabata and K. Binder, Solid State Commun. 
{\bf 22}, 705 (1977);
 S. Miyashita, H. Nishimori, 
A. Kuroda, M. Suzuki, Prog. of Theor. 
 Phys. {\bf 60}, 1669 (1978); 
 J. Tobochnik and G. V. Chester, 
Phys. \ Rev.\ B{\bf 20}, 
 3761 (1979);
 W. J. Shugard, J. D. Weeks and G. H. Gilmer, 
Phys. \ Rev. {\bf 21},  5309 (1980);
 J. van Himbergen, S. Chakravarty, 
Phys. \ Rev. B{\bf23}, 359 (1981);
 H. Betsuyaku, Physica {\bf106A}, 311 (1981);
 G. Fox, R. Gupta, O. Martin and S. Otto, 
Nucl. \ Phys. {\bf B205}, 188 (1982);
 M. Fukugita and Y. Oyanagi, 
Phys. \ Lett. {\bf B123}, 71 (1983); 
 S. Samuel, F. G. Yee, 
Nucl. \ Phys. {\bf B257}, 85 (1985);
 P. Harten, P. Suranyi, 
Nucl. \ Phys. {\bf B265}, 615 (1986);
 J. F. Fernandez, M. F. Ferreira  J. Stankiewicz, 
Phys. \ Rev. B{\bf 34}, 292 (1986);
H. Weber, P. Minnhagen, 
Phys. \ Rev. B{\bf37}, 5986 (1988);
P. Minnhagen and P. Olsson 
Phys. \ Scripta {\bf43}, 203 (1991).
\bibitem{ferervel}M. Ferer and M. J. Velgakis, 
Phys. \ Rev.\ B{\bf 27},  314 (1983).
\bibitem{seiler}E. Seiler, 
I. O. Stamatescu, A. Patrascioiu and V. Linke,  
 Nucl. \ Phys.[FS23] {\bf B305}, 623 (1988).
\bibitem{bcm} P. Butera, M. Comi and G. Marchesini, 
Phys.\ Rev.\ B {\bf 33}, 4725 (1986); 
P. Butera, R. Cabassi, M. Comi and G. Marchesini,
Comp.\ Phys.\ Comm. {\bf 44}, 143 (1987);
P. Butera, M. Comi and G. Marchesini, 
Phys.\ Rev.\ B {\bf 40}, 534 (1989).
\bibitem{bc} P. Butera, M. Comi, 
Phys.\ Rev.\ B {\bf 47},11969 (1993).
\bibitem{gupta} R. Gupta, J. De Lapp, 
G. G. Batrouni, G. C. Fox, C. F. Baillie and J. Apostolakis, 
Phys. \ Rev. \ Lett. {\bf91}, 1996 (1988); 
C. F. Baillie and R. Gupta, Nucl. \ Phys. B (Proc. Suppl.) 
{\bf 20}, 669 (1991);
 R. Gupta and C. F. Baillie , Phys.\ Rev.\ B {\bf 45}, 2883 (1992).
\bibitem{wolff} U. Wolff, Nucl.\ Phys. {\bf B322}, 759 (1989).
\bibitem{edwards}R. G. Edwards, J. Goodman, 
A. D. Sokal, Nucl.\ Phys. {\bf B354}, 289 (1991).
\bibitem{janke}W. Janke and  K. Nather, 
Phys. \ Lett. {\bf157A}, 11 (1991). 
\bibitem{biferale}L. Biferale, R. Petronzio, 
Nucl.\ Phys. {\bf B328}, 677 (1989).
\bibitem{sokal} 
J. Goodman, A. D. Sokal, 
Phys.\ Rev.\ Lett. {\bf 56}, 1015 (1986);
R. H. Swendsen, J. S. Wang, 
Phys. \ Rev. \ Lett. {\bf58}, 86 (1987);
M. Creutz, Phys.\ Rev.\ D {\bf 36}, 515 (1987); 
U. Wolff, Phys. \ Rev. \ Lett. {\bf62}, 361 (1989);
J. Goodman, A. D. Sokal, 
Phys.\ Rev.\ D {\bf 40}, 2035 (1989); 
A. D. Sokal, Nucl. \ Phys. B 
(Proc. Suppl.) {\bf 20}, 55 (1991).
\bibitem{kt} J. M. Kosterlitz and D. J. Thouless, 
J.\ Phys.\ {\bf C6}, 
 1181 (1973);
J. M. Kosterlitz, J.\ Phys.\ {\bf C7}, 1046 (1974).
\bibitem{bere} V. L. Berezinskii, 
ZETF {\bf 59}, 907 (1970)  (English trans.
Sov.\ Phys. JEPT {\bf 32}, 493 (1971) );
J. V. Jose, L. P. Kadanoff, 
S. Kirkpatrick and D. R. Nelson, Phys. Rev. 
B{\bf 16}, 1217 (1977); 
J. M. Kosterlitz and D. J. Thouless, in 
{\it Progress in low temperature physics },
Vol.VIIB, D. F. Brewer Ed. (Amsterdam , North Holland 1978);
A. P. Young, J. \ Phys. C{\bf11},  L453 (1978);
R. Savit, Phys. \ Rev. B{\bf17}, 1583 (1978);
J. B. Kogut,  Rev. \ Mod. \ Phys. {\bf 51}, 659 (1979);
AM. Steiner and A. R. Bishop , 
in {\it Solitons }, S. E. Trullinger, V. E. Zacharov,
V. L. Pokrovski Eds. (Berlin, Springer  1987);
P. Minnhagen, Rev.\ Mod.\ Phys. {\bf 59}, 1001 (1987);
J. Glimm and A. Jaffe, {\it Quantum Physics. 
 A functional integral point of view}, 
 (New York Springer Verlag  1987);
H. Kleinert, {\it Gauge fields in condensed matter}, Vol. I, 
 (Singapore, World Scientific 1989);
 D. P. Landau, J. Appl. Phys. {\bf 73}, 6091 (1993).
\bibitem {guerra}G. F. De Angelis, D. De Falco, F. Guerra,
\ Nuovo \ Cimento \ Lett.,{\bf 19},55(1977);  
F. Guerra, R. Marra, G. Immirzi,
\ Nuovo \ Cimento \ Lett., {\bf 23},237(1978); 
G. Marchesini, Nucl. Phys.{\bf B239}, (1984) 135; 
G. Marchesini, E. Onofri, Nucl. Phys.{\bf B249},  225 (1985); 
\bibitem{amit} P. B. Wiegman, 
J.\ Phys.\ C {\bf 11}, 1583 (1978);
D. J. Amit, Y.Y. Goldschimdt and G. Grinstein, 
J. Phys. A{\bf13},  585 (1980);
L. P. Kadanoff, A. B. Zisook, J. Phys. A{\bf13}, L379 (1980); 
L. P. Kadanoff, A. B. Zisook, 
Nucl. \ Phys. [FS2] {\bf B180},  61 (1981). 
\bibitem{aize} M. Aizenmann and B. Simon, 
Phys. \ Lett. {\bf76A}, 281 (1980).
\bibitem{tonyg} A. J. Guttmann in 
{\it Phase Transitions and Critical Phenomena},
C. Domb and J. Lebowitz Eds., 
Vol.13, ( London, Academic Press 1989); 
\bibitem{seran} G. A. Baker and D. L. Hunter, 
Phys. Rev. B{\bf7}, 3346, 3377(1973);
D. S. Gaunt, A.J. Guttmann in {\it Phase Transitions and 
Critical Phenomena},
C. Domb and M. S. Green Eds., 
Vol. 3 ,(London, Academic Press  1974);
C. J. Pearce, Adv. in Phys. {\bf27},1 (1978); 
G. A. Baker, 
{\it Quantitative Theory of Critical Phenomena}, London,
Academic Press 1990).
\bibitem{nickel} B.G. Nickel, in
 {\it Phase Transitions: Cargese 1980 },
M. Levy, J.C. Le Guillou and 
J. Zinn-Justin eds. ( New York, Plenum Press 
1982 )
\bibitem{diffapp} 
D. L. Hunter and G. A. Baker, 
Phys.\ Rev.\ B {\bf 19}, 3808 (1979); 
 M. E. Fisher and H. Au-Yang, 
J.\ Phys.\ A {\bf 12}, 1677 (1979);
J.J. Rehr , G, S. Joyce and A.J. Guttmann, 
J.\ Phys.\ A {\bf 13}, 1587 (1980);
B. G. Nickel and J. J. Rehr, J. Stat. Phys. {\bf 61}, 1 (1990).
\bibitem{sykes} M.F. Sykes, D. S. Gaunt, P. D. Roberts 
and J. A. Wyles, J.\ Phys.\ A {\bf 5}, 624 (1972);
\bibitem{tgut} A. J. Guttmann, J. \ Phys. \ A{\bf 20}, 1855 (1987). 
\end{references}
\end{document}